\documentclass[onecolumn]{openjournal}
\usepackage{natbib}
\usepackage{graphicx,amsmath,amssymb,amstext}
\usepackage{amsbsy,amsfonts,amsthm,color}

\usepackage[utf8]{inputenc}
\usepackage[caption=false]{subfig}
\providecommand{\del}[2]{\ensuremath{\frac{\partial #1}{\partial #2}}}
\usepackage[colorlinks,linkcolor=blue,citecolor=blue,urlcolor=blue ]{hyperref}

\begin{document}

\title{Classical Fluid Analogies for Schr\"odinger-Newton Systems}

\author{Peter Coles$^*$}
\email{$^*$Peter.Coles@mu.ie}
\author{Aoibhinn Gallagher}
\affiliation {Department of Physics, National University of Ireland, Maynooth, Co. Kildare, Ireland.}


\begin{abstract}
The Schr\"odinger-Poisson formalism has found a number of applications in cosmology, particularly in describing the growth by gravitational instability of large-scale structure in a universe dominated by ultra-light scalar particles. Here we investigate the extent to which the behaviour of this and the more general case of a Schr\"odinger-Newton system, can be described in terms of classical fluid concepts such as viscosity and pressure. We also explore whether such systems can be described by a pseudo-Reynolds number as for classical viscous fluids. The conclusion we reach is that this is indeed possible, but with important restrictions to ensure physical consistency.
\end{abstract}

\maketitle
\section{Introduction}
Recent years have seen an increasing interest in a particular formulation of the gravitational instability problem  - the Schr\"odinger-Poisson (SP) approach -  which is based on wave mechanics rather than fluid mechanics.  This approach was first proposed in the cosmological context by \cite{widrowKaiser1993}, and has subsequently generated various extensions and applications see e.g. \cite{coles2002wave}, \cite{Szap}, \cite{Coles_2003}, \cite{CS2006a}, \cite{CS2006b} and \cite{johnston2009cosmological}. It turns out that this approach, while initially developed as an approximate method for handling the evolution of cold dark matter, is also applicable to a scenarios in which the behaviour of dark matter is inherently wave-like rather than particle-like, such as if the dark matter is a very light bosonic particle, perhaps in the form of a condensate, or some other form of ``fuzzy'' dark matter; see for example \cite{Brook2022Gravitational}, \cite{Schwabe_2020}, \cite{https://doi.org/10.48550/arxiv.2111.01199},  \cite{hui2021wave} and references therein.

Most discussions of this approach have emphasized that the Schr\"{o}dinger equation of quantum mechanics can serve as an approximate representation of classical fluid flows. The reverse interpretation also holds, in that to some extent a classical fluid description can also serve to describe a quantum system. In this case, we take the Schr\"{o}dinger equation as our primary description and look for analogous fluid behaviour after doing the Madelung transformation in the opposite direction. This leads, for example, to a term often described as ``quantum pressure" in the fluid equations.

One of the aims of this paper is to follow on from the analysis in \cite{Gallagher2022Evolution}  by discussing the extent to which this ``quantum pressure'' can indeed be interpreted as a pressure in a true analogy with a classical fluid. Pressure in fluids is typically split into two categories, hydrostatic and dynamic. Hydrostatic pressure refers  to the pressure of a fluid that is not moving; dynamic pressure refers to pressure of a moving fluid in a closed system. It seems reasonable to treat cosmological fluids in a similar way to the ocean, i.e. as a hydrostatic case, in which the pressure changes due to  the motion of the fluid are negligible. Pressure within fluids can be calculated in many situations by the Bernoulli equation,
but the derivation of this equation requires the fluid to have zero viscosity. So, does this apply? It has also been remarked previously that the coefficient $\nu$ we have introduced has the same dimensions as a kinematic viscosity; see, for example, \cite{CS2006a, CS2006b} for discussion. 

\section{Background Cosmology}
\label{cosmo} 
The current standard model of cosmology assumes the Cosmological Principle, according to which the universe is assumed to be homogeneous and isotropic, at least on large scales. Space-times consistent with this are described by the Robertson-Walker metric: 
\begin{equation}
    d s^2 = c^2 d t^2 - a^2(t)\left( \frac{d r^2}{1-\kappa r^2} + r^2d \theta^2 +r^2 \sin^2(\theta) d \phi^2 \right),\label{eq:RWmetric}
\end{equation}
where $\kappa$ is the spatial curvature: $\kappa=0$ represents flat space; $\kappa = +1$ represents constant positively curved space (closed universe); and $\kappa = -1$ represents constant negatively curved space (open universe). The time coordinate $t$ is cosmological proper time and $a(t)$ is the cosmic scale factor. The time evolution of the cosmic scale factor, $a(t)$, is determined via Einstein's gravitational field equations through the Friedman equation,
\begin{equation}
    3\left( \frac{\dot{a}}{a} \right)^2 = 8\pi G \rho - \frac{3\kappa c^2}{a^2} + \Lambda \label{eq:Freidman}
\end{equation}
the deceleration equation,
\begin{equation}
    \frac{\ddot{a}}{a} = -\frac{4 \pi G}{3}\left( \rho + 3\frac{p}{c^2} \right) + \frac{\Lambda}{3} \label{eq:deceleration}
\end{equation}
and the density-pressure relation,
\begin{equation}
    \dot{\rho} = -3\frac{\dot{a}}{a}\left( \rho + \frac{p}{c^2} \right) \label{eq:einstein3}
\end{equation}
where the dots denote derivatives with respect to cosmological proper time t, thus describing the global expansion or contraction of the universe. These models can be further parametrised by the Hubble parameter $H =\dot{a}/a$ and the density parameter $\Omega = 8\pi G \rho/3 H^2$. As usual the present epoch is defined by $t = t_0$, when $H = H_0$ and $\Omega = \Omega_0$. 

Throughout this paper we assume Newtonian gravity, since the scale of the perturbations are much smaller than the effective cosmological horizon $d_H = c/H$. We also do all calculations in co-moving coordinates, 
\begin{equation}
    \textbf{x} \equiv \textbf{r}/ a(t), 
    \label{eq:comoving}
\end{equation}
where $\textbf{x}$ is the co-moving spatial coordinate ($\textbf{x}$ is fixed in the frame of the Hubble expansion), $\textbf{r}$ is the spatial coordinate in our reference frame, and $a(t)$ is the cosmic scale factor.

We start with the Newtonian equations for a self-gravitating perfect fluid: the \textit{Euler Equation}
\begin{equation}
    \del{\textbf{v}}{t} + (\textbf{v}\cdot \nabla ) \textbf{v} + \frac{1}{\rho} \nabla p + \nabla V = 0 \label{eq:euler};
\end{equation}
the \textit{Continuity Equation}
\begin{equation}
    \del{\rho}{t} + \nabla \cdot(\rho \textbf{v}) = 0 \label{eq:continuity};
\end{equation}
and the \textit{Poisson Equation}
\begin{equation}
    \nabla^2 V = 4\pi G \rho
    \label{eq:poisson}.
\end{equation}
Here ${\bf v}$, $p$ and $\rho$ are the fluid velocity, pressure and density, respectively, and $V$ here is the gravitational potential.

The linear evolution of small perturbations is a well-studied problem that can be solved analytically for any background cosmological model. In the non-linear regime, however, analytical results are hard to come by and we are usually forced to rely on numerical calculations.

A simple and elegant analytical approach to the non-linear stage of gravitational evolution is the Zel'dovich approximation, introduced by \cite{zeldovich}.  This is an excellent tool for studying structure formation and is highly accurate up to the point of shell crossing \cite{osti_6192112}, at which point it breaks down. To combat this we need a model that goes beyond shell crossing, this is one reason for interest in the SP model. Other methods for studying multi-stream flow can be seen in \cite{multistream} and \cite{Gough2022}. 

A different approach to this problem can be found in the  so-called adhesion approximation; see \cite{GSS89}, an extension of the Zel'dovich approximation achieved by introducing an effective viscosity term to make particles ``stick'' at shell-crossing. A further extension of this model is presented by \cite{Jones1999}, analysed from a wave--mechanical point of view in \cite{Coles2002}, and discussed further below.

\section{The Schr\"odinger-Poisson Description}
\label{sp system} 
 We now simplify our treatment by assuming a cold irrotational fluid moving under the influence of a general potential $V$. This means that Eq. (\ref{eq:euler}) now becomes the much simpler \textit{Bernoulli equation}
\begin{equation}
    \del{ \phi}{t} + \frac{1}{2}(\nabla \phi)^2 = -V \label{eq:bernoulli}, 
\end{equation}
where $ \textbf{v} = \nabla \phi, $ where $\phi$ is the velocity potential. 
We now make the \textit{Madelung Transformation}
\\
$\psi = \alpha e^{i\phi/\nu} $, where $\rho = \psi \psi^* = \alpha^2$. This gives 
\begin{equation}
    i \nu \del{\psi}{t} = -\frac{\nu^2}{2} \nabla^2 \psi + V \psi + P \psi \label{eq:schrodgen}
\end{equation}
where 
\begin{equation}
    P = \frac{\nu^2}{2} \frac{\nabla^2 \alpha}{\alpha}, \label{eq:QMpressure}
\end{equation}
from Equations (\ref{eq:bernoulli}) and (\ref{eq:continuity}). 
Equation (\ref{eq:schrodgen}) is  Schr\"odinger Equation, with potential $V$, $\nu$ acting as $\hbar$ with the addition of non-linear term $P$. Although $\psi$ is governed by the same equation as the evolution of a single-particle wave function, that is not how it should be interpreted. In particular, $|\psi|^2$ represents a physical density not a probability density, and its evolution is completely unitary - there is nothing like the wave-function collapse that occurs in standard quantum mechanics in this system.

It's important to note a crucial advantage of this description, namely that because $\rho=|\psi|^2$ the condition that $\rho\geq 0$ is automatically enforced if one applies, e.g., perturbation theory to $\psi$. This is not the case for approaches based on standard Eulerian perturbation theory applied to $\delta=(\rho-\rho_0)/\rho_0$ which predict $\rho<0$ when $\delta<-1$ and are therefore very unsuitable for describing voids. Note also that because the wave-function describes a delocalized particle there are no singularities analogous to the caustics that form in the Zel'dovich approximation.

In the cosmological context we take $V$ to be the gravitational potential determined via the Poisson Equation and we get a system of coupled Partial Differential Equations (PDEs). We set $\nu$ to be an adjustable parameter so we have some control over the frequency of any oscillatory solutions that could arise from this system. $\nu$ has dimensions such that $\phi / \nu$ is dimensionless, $\phi$ is a velocity potential so has dimensions $L^{2} T^{-1}$, these are also the dimensions of kinematic viscosity, so $\nu$ can be thought of as a viscosity parameter for the model; we return to this later.

The original intentions of this model were to find a fluid interpretation of a quantum system, \cite{widrowKaiser1993} first proposed this be used to simulate cold dark matter (CDM), and it has been used predominantly in a cosmological context since. The quantum nature of this model is also an advantage as it gives us a new perspective on the behaviour of large scale structure. This is especially useful when one seeks to look beyond shell crossing, as this model can handle multiple streams. 

Assuming a potential flow for the velocity field ${\bf u}$ such that ${\bf u}=-{\bf\nabla}\phi$, the adhesion approximation boils down to the following equation for $\phi$:
\begin{equation}
\frac{\partial \phi}{\partial \tau} - \frac{1}{2}|{\bf\nabla}\phi|^2 - \mu \nabla^2\phi= 0,
\end{equation}
where $\mu$ (which is assumed constant) is the viscosity.   \cite{CS2006a} showed that in the wave-mechanical representation the corresponding form is:
\begin{equation}
\frac{\partial \phi}{\partial \tau} - \frac{1}{2}|{\bf\nabla}\phi|^2 + {\cal P}=0,
\end{equation}
where ${\cal P}$ is the quantum pressure:
\begin{equation}
{\cal P}=\frac{\nu^2}{2} \frac{\nabla^2 |\psi|}{|\psi|}.
\end{equation}
The $\nu$-term plays an important role in the dynamics of this system just as the $\mu$-term does in the former case, but it has quite a different form of spatial variation. Further connections between the adhesion model and the Schr\"{o}dinger approach are explored in \cite{rib2005}.

The question thus arises as to the extent that we can treat $\nu$ as a term describing viscosity in a quantum system. The viscosity of a fluid is its ability to resist change in shape, so it quantifies the internal friction of the fluid; see \cite{stokes1849}. Viscous fluids are often associated with vorticity, and our representation requires a potential flow for which the vorticity is zero by definition. On the other hand, there is a sizeable literature \cite{JOSEPH2006} on the subject of potential flows in viscous fluids so we feel it is worth taking this discussion further, though we are obviously limited in what we can say without expanding the description of velocity fields beyond the current simple form. An example of an application of this general approach to a fluid with viscosity is given by  \cite{Brook2022Gravitational}.

\section{Viscosity in a Navier-Stokes Equation}
First, start with the full Navier-Stokes equation in the form
\begin{equation}
\frac{\partial {\bf v}}{\partial t} +({\bf v}\cdot{\bf\nabla}) {\bf v} + \frac{{\bf\nabla}P}{\rho} - \frac{\eta}{\rho} \nabla^2 {\bf v} - \frac{1}{\rho} \left(\zeta + \frac{\eta}{3} \right) {\bf\nabla} \left({\bf\nabla} \cdot {\bf v}\right) = {\bf0}, \label{eq:NSFull}
\end{equation}
where we have used the decomposition of the viscous stress tensor $\sigma_{ij}$:
\begin{equation}
\sigma_{ij} = \eta \left(\frac{\partial v_i}{\partial x_j} + \frac{\partial v_j}{\partial x_i} - \frac{2}{3} \delta_{ij} \frac{\partial v_k}{\partial x_k} \right) + \zeta \delta_{ij} \frac{\partial v_k}{\partial x_k},
\end{equation}
in which $\zeta$ is the bulk viscosity; $\eta$ is the shear viscosity and $v_i$ are components of the velocity field. We know that our approach only works for a potential flow so we specialize to the
case of an irrotational flow, which means we can drop the last term on the LHS of Eq. (\ref{eq:NSFull}).

This means we can write the Navier-Stokes equation in the form
\begin{equation}
\frac{\partial {\bf v}}{\partial t} +({\bf v}\cdot{\bf \nabla}) {\bf v} + \frac{{\bf \nabla}P}{\rho} - \frac{\eta}{\rho} \nabla^2 {\bf v}={\bf 0}.
\end{equation}
In this and the following $\eta$ is the effective viscosity coefficient which is obtained by redefinition of the original $\eta$ to absorb a contribution from $\zeta$:
\begin{equation}
\eta \rightarrow \zeta + \frac{4\eta}{3}.
\end{equation}

If we neglect the pressure (e.g. for a cold fluid) then the Navier-Stokes equation then becomes
\begin{equation}
\frac{\partial {\bf v}}{\partial t} +({\bf v}\cdot{\bf \nabla}) {\bf v}  - \frac{\eta}{\rho} \nabla^2 {\bf v}={\bf 0}. \label{eq:NSnoP}
\end{equation}

The first point to be made is that there is little point in talking about fluid analogies without including all the quantities needed for a full classical description. If we write the Navier-Stokes equation in terms of the material derivative, it becomes
\begin{equation}
\rho\frac{D{\bf v}}{Dt} = -{\bf \nabla}P+{\bf \nabla}\cdot{{\bf \sigma}},
\end{equation}
in which the final term represents the effects of internal friction. It is therefore inconsistent to describe this system without including a description of heat transfer - including viscous dissipation and associated as entropy production which would require the introduction of a temperature $T$ and entropy $s$ as well as thermal conductivity $\kappa$. The general form of this would be
\begin{equation}
\rho T \frac{Ds}{Dt} = {\bf \nabla}\cdot{\kappa {\bf \nabla}T} + \frac{\eta}{2} \left(\frac{\partial v_i}{\partial x_j} + \frac{\partial v_j}{dx_i}- \frac{2}{3} \delta_{ij} {\bf \nabla}\cdot{\bf v}\right)^2 + \zeta ({\bf \nabla}\cdot{\bf v})^2,
\end{equation}
in which the first term on the right-hand side represents the thermal conduction and the other two describe viscous dissipation. 
To set up a full fluid analogy would therefore involve complexities beyond the scope of this work. What we will do, however, is see how far we can get by starting with the Schr\"{o}dinger equation. In the following we are guided by ref. \cite{SvsNS2016} to whom we refer the reader for further comments.

Since we are interested in the description of a quantum system (in the form of a light scalar particle) we will not initially use the parameter $\nu$ employed elsewhere but retain the factors of $\hbar$, $m$, etc. Consider the Schr\"{o}dinger equation for a particle of mass $m$ in a time-independent potential $V({\bf x})$, which can be written
\begin{equation}
i\hbar \frac{\partial \psi}{\partial t} + \frac{\hbar^2}{2m} \nabla^2 \psi - V\psi=0.
\end{equation}
We make the usual Madelung transformation written in the form\begin{equation}
\psi=\psi_0 \exp\left(S+i\phi\right) =  \psi_0 N \exp(i\phi),\end{equation}
in which $\psi_0$ is a normalizing constant that can be included in $R$ but it is useful in describing the Boltzmann entropy:
\begin{equation}
{\cal S}= k_B\log_e \left[|\frac{\psi}{\psi_0}|^2 \right],\end{equation}
which is related to $S$.

With this choice of definitions\begin{equation}
\rho = m |\psi|^2.\end{equation}
There is a length scale $\lambda$ in $\psi$ defined by the normalization factor $\psi_0$ because we have to integrate $\psi^2$ over a volume to get unity:
\begin{equation}
\lambda= |\psi_0|^{-2/3},
\end{equation}
which means that 
\begin{equation}
\rho = m |\psi|^2 = \frac{m}{\lambda^3}\exp(2S) = \frac{m}{\lambda^3} N^2.
\end{equation}
Now the Madelung transformation leads to two equations:
\begin{equation}
\frac{\partial S}{\partial t} + \frac{\hbar}{m} {\bf \nabla}S \cdot {\bf \nabla}\phi + 
\frac{\hbar^2}{2m} \nabla^2 \phi = 0;
\end{equation}
and
\begin{equation}
\hbar \frac{\partial \phi}{\partial t} + \frac{\hbar^2}{2m} (\nabla \phi)^2 + V+ U=0, \label{eq:Schrod_full}
\end{equation}
where
\begin{equation}
U= - \frac{\hbar^2}{2m} \frac{\nabla^2 N}{N} = - \frac{\hbar^2}{2m} \left[(\nabla S)^2 + \nabla^2 S\right],
\end{equation}
which we might term the quantum potential. In the general case $V$ could be an external potential or, as in the specific cases we have discussed in this thesis, a potential resulting from self-gravitational interactions. Taking the gradient of
Eq. (\ref{eq:Schrod_full}) and using the velocity potential $\phi$ to determine the velocity field ${\bf v}$,
\begin{equation}
{\bf v}=\frac{\hbar}{m} {\bf \nabla} \phi,
\end{equation}
to get
\begin{equation}
    \frac{\partial {\bf v}}{\partial t} +({\bf v}\cdot{\bf \nabla}) {\bf v}  + \frac{1}{m} {\bf \nabla}U +\frac{1}{m} {\bf \nabla}V ={\bf 0}. \label{eq:NSSchrod}
\end{equation}

\cite{SvsNS2016} argue that, in order to make this equation compatible with  Eq. (\ref{eq:NSnoP}),  it is necessary to impose two separate conditions: first that
\begin{equation}
\frac{{\bf \nabla} P}{\rho}= \frac{{\bf \nabla}V}{m},;\end{equation}
which is tantamount to the assumption that we are dealing with hydrostatic pressure and also means that if $V=0$ we must use $P=0$ for consistency; and second we have to
identify the term in $U$ with the viscosity term, i.e.\begin{equation}
\frac{1}{m} {\bf \nabla}U= -\frac{\eta}{\rho} \nabla^2 {\bf v}. \label{eq:identify}
\end{equation}
This condition means that the gradient of the quantum potential must exactly balance the viscosity term in the equations of motion. Internal frictional forces arising in a system described by these equations are therefore inherently quantum-mechanical in nature. Eq. (\ref{eq:identify}) gives
\begin{equation}
\frac{\eta}{\rho} \nabla^2 {\bf v}= - \frac{1}{m} {\bf\nabla} U = - \frac{1}{m}
{\bf\nabla}\left(\frac{-\hbar^2}{2m} \frac{\nabla^2 N}{N} \right),
\end{equation}
with 
\begin{equation}
{\bf v}=\frac{\hbar}{m} {\bf\nabla \phi}.\end{equation}
This means that the viscosity term is in general spatially dependent and very complicated but is indeed of order $\hbar$ which is the salient point. If we want to describe this system using the Navier-Stokes equations (which have constant $\eta$) we have to take the semi-classical limit in which $\hbar \rightarrow 0$ and that the spatial variations of the wave-function are small. This in turn means that $\nabla^2\phi\simeq 0$ which it is easy to show requires that the fluid be almost incompressible and the flow be almost isentropic (i.e. there is negligible entropy generation).

One could go further by analysing the behaviour found in \cite{johnston2009cosmological} and \cite{Coles_2003}. Then we can replace
the partial derivatives by ordinary derivatives, assume ${\bf v}$ only has an $x$-component called $v$ and write
\begin{equation}
\eta(x)=\rho(x) \frac{A(x)}{B(x)},\label{eq:rat}
\end{equation}
where
\begin{equation}
\rho(x)= \frac{m}{\lambda^3} N^2
\end{equation}
\begin{equation}
A(x)= -\frac{\hbar^2}{2m^2}\frac{d}{dx}\left(\frac{\frac{d^2 N}{dx^2}}{N} \right)= -\frac{\hbar^2}{2m^2}\frac{d}{dx} \left[ \left(\frac{dS}{dx}\right)^2 +  \frac{d^2 S}{dx^2}\right]\end{equation}
and
\begin{equation}
B(x)= \frac{d^2 v}{dx^2} = \frac{d^2}{dx^2} \left( \frac{\hbar}{m} \frac{d\phi}{dx} \right) = \frac{\hbar}{m} \frac{d^3\phi}{dx^3}.\end{equation}
Two things one can see immediately from these is that $m$ cancels on the right-hand side of Eq. (\ref{eq:rat}) and the result for $\eta$ is of order $\hbar$ as argued above. For qualitative understanding, therefore, there is therefore no need to include factors of $\lambda$, $m$ or $\hbar$ in any calculation. We just need 
\begin{equation}
|\eta(x)| = \frac{N^2 \frac{d}{dx}\left(\frac{\frac{d^2 N}{dx^2}}{N} \right)}{\frac{d^3\phi}{dx^3}} =
\frac{N\frac{d^3N}{dx^3} - \frac{d^2 N}{dx^2}\frac{dN}{dx}}{\frac{d^3\phi}{dx^3}}.
\end{equation}
Assuming a given form for $N(x)$ and $\phi(x)$ would straightforwardly give an effective value behaviour of $\eta(x)$ for a given configuration if one were interested in doing so. The main point of this section however is that the description of this system in terms of a Navier-Stokes equation is only consistent in the limit $\hbar\rightarrow 0$ in which case $\eta\rightarrow \infty$. That case is not irrelevant: it is, for example, the limit taken in analytic treatments of the adhesion model.

\section{Example: Wave Propagation}

Having established the possibilities and limitations of the incorporation of viscosity into the description of quantum fluids it makes sense to consider a physically motivated example that can be done consistently.  Given that we can not describe vorticity we have a relatively restricted set of examples to choose from, but one that is relevant is the propagation of internal waves within such a fluid. The internal friction as a consequence of having a viscosity, causes dissipation of energy and dampens waves as they travel through the fluid. This results in attenuation of the wave amplitude over distance. For consistency with the considerations of the previous section, we treat small-amplitude waves only.

To determine an attenuation of ``acoustic'' waves we begin with conservation of both mass and momentum.  Conservation of mass is given by the \textit{Continuity Equation}, 
\begin{equation}
    \del{\rho}{t} + \nabla \cdot (\rho {\bf v}) = 0,
    \label{eq:continuity3}
\end{equation}
and conservation of momentum is given by
\begin{equation}
    \rho \frac{D{\bf v}}{Dt} + \nabla {\bf P} = \frac{4 \mu}{3} \nabla ^{2} {\bf v}.
    \label{eq:cons_mom}
\end{equation}
We define the \textit{material derivative} for velocities in a fluid to be 
\begin{equation}
    \frac{D}{Dt} \equiv \del{}{t} + {\bf v} \cdot \del{}{x}.
    \label{eq:material_deriv}
\end{equation}

The pressure ${\bf P} = (P, P, P)$ is given as a vector for consistency. The \textit{dynamic viscosity} is included 
in Eq. (\ref{eq:material_deriv}) as $\mu$. The dynamical viscosity relates to the shear or kinematic viscosity by $\mu = \nu \rho$. 

For fluids, the equation of state relates pressure, density and entropy, by expressing pressure as a function of  density and entropy. For cosmic fluids we have the pressure as a function of density in the form of the  quantum pressure, 
\begin{equation}
    P = \frac{\nu^2}{2} \frac{\nabla^{2}(\sqrt{\rho})}{\sqrt{\rho}}.
    \label{eq:pressure}
\end{equation}
To move forward with this as our equation of state, we will use the more convenient form 
\begin{equation}
    P = \nu^2 \left[ \frac{\nabla ^{2}\rho}{2 \rho} - \biggl( \frac{\nabla \rho}{2 \rho} \biggr)^2 \right]
    \label{eq:pressure_new}
\end{equation}

For convenience and brevity we will use notation in which the subscript denotes derivative, for example,  
\begin{equation}
    \rho_t = \del{\rho}{t}.
\end{equation}
Sound waves typically disturb the fluid only around a small region, therefore, the 
quantities associated with sound, excess pressure, excess density and particle velocity 
can be assumed to be small and of first order. Assume
\begin{equation}
    | \delta \rho | \ll \rho_0.
    \label{eq:smsig}
\end{equation}
The Continuity Equation (\ref{eq:continuity3}) can be expanded to the following, 
\begin{equation}
    \delta \rho_t + v \delta \rho_x + v_x \rho_0 + v_x \delta \rho = 0.
    \label{eq:continuity_expanded}
\end{equation}
It can be seen that this reduces to
\begin{equation}
    \delta \rho_t + v_x \rho_0 = 0,
    \label{eq:cont_smsig}
\end{equation}
keeping only first-order terms. Following the same procedure, we can see that Eq. (\ref{eq:cons_mom}) becomes
\begin{equation}
    \rho_0 v_t + P_x = \frac{4 \nu}{3} \rho_0 v_{xx},
    \label{eq:cons_mom_smsig}
\end{equation}
and Eq. (\ref{eq:pressure_new}) becomes 
\begin{equation}
    P = \frac{\nu^2}{4} \frac{\delta \rho_{xx}}{\rho_0 + \delta \rho}.
    \label{eq:pressure_smsig}
\end{equation}
The small signal approximation is encapsulated by
Eqs (\ref{eq:cont_smsig}), (\ref{eq:cons_mom_smsig}) and (\ref{eq:pressure_smsig}). Combining
these equations produces a final form of the viscous wave equation. From Eq. (\ref{eq:cons_mom_smsig}) we get
\begin{equation}
    \begin{split}
    \rho_0 v_x &= -\delta \rho_t, \\
    \rho_0 v_{xx} &= -\delta \rho_{tx}, \\
    \rho_0 v_{tx} &= -\delta \rho_{tt}, \\
    \rho_0 v_{xxx} &= -\delta \rho_{txx},
    \end{split}
    \label{eq:diffs}
\end{equation}
Differentiate Eq. (\ref{eq:cons_mom_smsig}) with respect to $x$ and use identities 
found in Eq. (\ref{eq:diffs}) to get, 
\begin{equation}
    - \delta \rho_{tt} + P_{xx} = - \frac{4 \nu}{3} \delta \rho_{txx}.
    \label{eq:cons_mom_diff1}
\end{equation}
Differentiating Eq. (\ref{eq:pressure_smsig}) with respect to $t$ and discarding higher order terms we get,
\begin{equation}
    P_t = \frac{\nu^2}{4} \frac{\delta \rho_{txx}}{\rho_0 + 2 \delta \rho}.
    \label{eq:pressure_diff}
\end{equation}
Rearranging gives 
\begin{equation}
    \delta \rho_{txx} = \frac{4}{\nu^2} \rho_0 P_t .
    \label{eq:delta_rho_txx}
\end{equation}
Substituting into Eq. (\ref{eq:cons_mom_diff1}) gives,
\begin{equation}
    - \delta \rho_{tt} + P_{xx} = -\frac{16}{3 \nu} \rho_0 P_t.
\end{equation}
Then, differentiating with respect to $x$ twice gives,
\begin{equation}
    - \delta \rho_{ttxx} + P_{xxxx} = -\frac{16}{3 \nu} \rho_{0} P_{txx}.
    \label{eq:cons_mom_diff2}
\end{equation}
Differentiating Eq. (\ref{eq:delta_rho_txx}) with respect to $t$, we find
\begin{equation}
    \delta \rho_{ttxx} = \frac{4}{\nu^2} \rho_0 P_{tt} .
    \label{eq:delta_rho_ttxx}
\end{equation}
Combining Eq. (\ref{eq:cons_mom_diff2}) and Eq. (\ref{eq:delta_rho_ttxx})
gives the final form of the viscous wave equation
\begin{equation}
    \frac{\nu ^2}{4} P_{xxxx} - \rho_0 P_{tt} - \frac{4 \nu}{3} \rho_0 P_{txx}.
    \label{eq:viscous_pressure}
\end{equation}

To solve Eq. (\ref{eq:viscous_pressure}) it is fair to assume a general wave  solution and look for a particular solution.  Assume that
\begin{equation}
    P = P_0 e^{i (\omega t - kx)}.
    \label{eq:p_solution}
\end{equation}
We seek a dispersion relation $(k = k(\omega))$ of the form, 
\begin{equation}
    k = \beta - i \alpha ,
    \label{eq:k_assume}
\end{equation}
where $\alpha$ will be the attenuation coefficient, and $\beta$ will define the phase velocity $c = \omega / \beta$.

Solving Eq. (\ref{eq:viscous_pressure}) with Eq. (\ref{eq:p_solution}) for $k(\omega)$ 
gives

\begin{equation}
    k = \pm \frac{2 \sqrt{\rho_0} \omega \sqrt{\frac{\nu^2}{4} + 
    i \frac{4 \nu}{3} \rho_0 \omega}}{\sqrt{\frac{\nu^4}{16} + 
    \frac{16 \nu^2}{9} \rho^2_0 \omega ^2}}.
    \label{eq:k_solution}
\end{equation}
Clearly, this is a complicated expression from which $\alpha$ is not easily extracted. 
However, the point is that $\alpha$ exists and can be found for specified initial  conditions. 

The attenuation of sound waves in an ``Fuzzy Dark Matter'' (FDM) fluid corresponds to similar damping effects in, for example, in Hot Dark Matter (HDM) caused by free-streaming in that it would suppress the formation of small-scales structures, 
relative to the predictions of CDM in which free streaming is negligible. 

\section{A pseudo-Reynolds Number}
\label{viscosity:reynolds}

To bring this discussion of viscosity full circle we discuss the applicability of a dimensionless scaling Reynolds number representing the ratio of inertial to viscous forces within a fluid. In
\cite{Gallagher2022Evolution}, we defined the Reynolds number (introduced by Stokes (\cite{stokes1851}), popularised by Reynolds (\cite{reynolds1883}), but named by Sommerfeld (\cite{sommerfeld1908}) as
\begin{equation}
    R = \frac{u l}{\nu}.
    \label{eq:reynolds_number}
\end{equation}
Here $u$ is a velocity, $l$ a length scale, and $\nu$ the viscosity (to go back to our earlier notation). All three components
are chosen to be reasonably associated with the motion. Although $\nu$ is now well 
defined, it is not clear how one would define the other two quantities. There are 
many velocity quantities present in this model and it is not exactly clear which one we 
should choose. There is a momentum associated with the wave-function, but this is not 
uniform in space and it is not clear how one could convert this to a single number.  There is also the issue of a length scale associated with the motion. It may be obvious that we choose to 
use the size of the object we are studying, however, in dealing with both collapse and void expansion this 
length is constantly changing. This length is also only unique in one-dimension, but we will stick to that case for this discussion.

\begin{figure}[htbp]
    \centering 
    \includegraphics*[width = 0.6\linewidth]{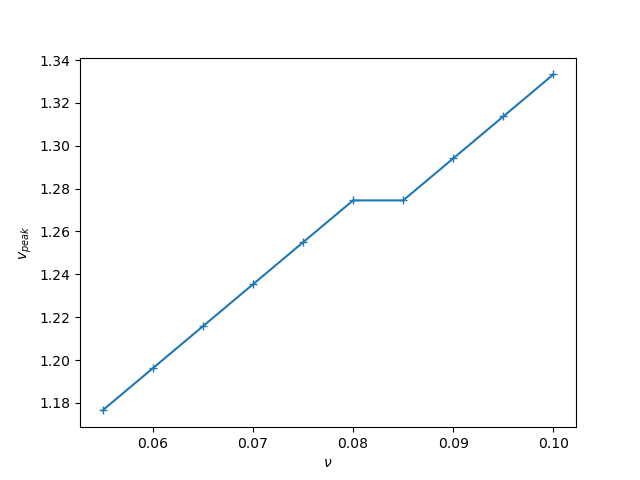}
    \caption{Peak velocity of a one dimensional SP void  plotted against viscosity
    parameter $\nu$}
    \label{fig:nu_v_corr}
\end{figure}

\cite{Gallagher2022Evolution} used the expansion of cosmological voids as a test for this idea. For the velocity, we used the expansion speed for the void; under-dense regions expand more quickly than the background cosmology and generate a clear edge whose motion we could use as a fiducial speed which we call $v_{\mathrm peak}$. With all of the parameters defined in a systematically consistent way for void expansion,  we could test the reliability of a pseudo-Reynolds number by exploring the relationships 
between the quantities used to define it. Since $u$ is the calculated value $v_{\mathrm peak}$,  we started by fixing $l$, and comparing various $\nu$ values with the $v_{\mathrm{peak}}$ 
produced by this. As seen in Figure (\ref{fig:nu_v_corr}) it is clear that $\nu$ and $v_{\mathrm{peak}}$ are indeed roughly proportional to each other. 
(The ''glitch'' is probably caused by the formation of a secondary peak that temporarily introduces a different velocity into the problem). Similar scaling can be verified for fixed $\nu$ and  calculated $v_{\rm peak}$ for varying $l$ values  and  at fixed $v_{\mathrm peak}$  between $\nu$ and $l$.
In summary, \cite{Gallagher2022Evolution}) found that each parameter has the correct relationship to 
infer a Reynolds number, so scaling solutions of a sort do exist in this system.   With a Reynolds number defined for one-dimensional collapse and one-dimensional void expansion, this is sufficient proof of concept for a pseudo-Reynolds number of in this particular Schr\"{o}dinger-Poisson system.

\section{Conclusions}
In this article, we discussed the description of pressure and viscosity in the context of a quantum fluid described by a Schr\"{o}dinger equation, coupled to an equation describing Newtonian interactions, in the archetypical case Poisson's equation for gravitation. In particular, we explored the possibility of interpreting the dynamics in terms of the Navier-Stokes equation rather than the usual Euler equation, finding that it is possible but only in restricted circumstances. We went on to illustrate this approach by studying the attenuation of acoustic waves in such a fluid. Finally, we justified the use of an effective viscosity parameter to construct a pseudo-Reynolds number that can be used as a scaling parameter for such systems, as in the case of classical fluid dynamical problems. This latter observation may be useful in more complex settings of the Schr\"{o}dinger-Newton or Schr\"{o}dinger-Poisson system. A detailed exploration of the use of this sort of scaling in simulations of ultra-light dark matter would be an interesting topic for future work.

\section*{Acknowledgments}
We thank Cora Ulhlemann, Alex Gough and Paul Watts for helpful discussions. We also acknowledge use of the NASA/ADS system in preparing this paper.

\newpage
\bibliographystyle{mnras}
\bibliography{bibliography}

\end{document}